# Onset of nonlinear electroosmotic flow under AC electric field

Zhongyan Hu,[1] Wenxuan Zhao,[1] Yu Chen,[1] Chen Zhang,[1] Xiaoqiang Feng,[1] Guangyin Jing,[2] Kaige Wang,[1] Jintao Bai,[1] Guiren Wang [3] and Wei Zhao [1,*]

[1] State Key Laboratory of Photoelectric Technology and Functional Materials, International Scientific and Technological Cooperation Base of Photoelectric Technology and Functional Materials and Application, Institute of Photonics and Photon-technology, Northwest University, Xi'an 710127, China
[2] School of Physics, Northwest University, Xi'an 710127, China
[3] Department of Mechanical Engineering & Biomedical Engineering Program, University of South Carolina, Columbia SC, 29208, USA.
* Corresponding author: zwbayern@nwu.edu.cn

**Nonlinearity of electroosmotic flows (EOFs) is ubiquitous and plays a crucial role in the mass and energy transfer in ion transport, specimen mixing, electrochemistry reaction, and electric energy storage and utilizing. When and how the transition from a linear regime to a nonlinear one is essential for understanding, prohibiting or utilizing nonlinear EOF. However, suffers the lacking of reliable experimental instruments with high spatial and temporal resolutions, the investigation of the onset of nonlinear EOF still stays in theory. Herein, we experimentally studied the velocity fluctuations of EOFs driven by AC electric field via ultra-sensitive fluorescent blinking tricks. The linear and nonlinear AC EOFs are successfully identified from both the time trace and energy spectra of velocity fluctuations. The critical electric field ($E_{A,C}$) separating the two statuses is determined and is discovered by defining a generalized scaling law with respect to the convection velocity ($U$) and AC frequency ($f_f$) as $E_{A,C} \sim f_f^{0.48-0.027U}$. The universal control parameters are determined with surprising accuracy for governing the status of AC EOFs. We hope the current investigation could be essential in the development of both theory and applications of nonlinear EOF.**

Electric double layer (EDL), is widely present in nature and engineering, directly resulting from the inhomogeneous distribution of ions and counterions on two-phase interfaces down to nanometer scale. For a long period, understanding the dynamics of EDL and its consequence behavior under an external electric field, i.e. electroosmotic flow (EOF), is crucial for electrochemistry, micro/nanofluidics, energy and environmental disciplines. In a complete electrochemistry system, normally both electrode/liquid and insulator/liquid interfaces coexist. Under an alternating current (AC) electric field, the ions in EDL are oscillating by the driven of electrostatic force, and induce AC EOF on both the electrode/liquid and insulator/liquid interfaces, which are widely adopted in biosensor chips, DNA sequencing, hybridization, micro and nanofluidic transportation[1], electrochemical reactions[2], liquid-based supercapacitors[3] and electrolyte batteries[4] where fast charge-discharge cycles commonly exist.

EOF, as the straightforward process in an electrochemistry system, is focusing on the ultrathin electrode/liquid interface (as shown in Fig. 1a which is the simplest model of an electrochemistry system). Usually, when the external electric field intensity is low, the EDL on the electrode/liquid interface remain in equilibrium, and the oscillating EOF on the insulator/liquid interface oscillates linearly to the external electric field, as comprehensively studied by both numerical simulations and experiments on both Newtonian and non-Newtonian fluids.[5-9] While the electric field is sufficiently high, the EDL on the electrode/liquid interface can be far from the equilibrium state, where transient concentration polarization near the electrode becomes dominant. It subsequently leads to a nonlinear internal electric field and nonlinear oscillating EOF on the insulator/liquid interface with respect to the external electric field. The periodic concentration polarization generated by AC electric field, especially the electric potential and ion flux, has attracted more attention from theoretical researchers[10,11] in the last two decades. By the conventional flow visualization [12] and particle tracking [13], it is possible already to reveal the strong nonlinear flow due to highly depleted ion concentration. However, all the current experimental investigations[14-19] are suffering the low sensitivity being able to address the even basic and two key issues, which are: (1) what parameters control the transition of AC EOF and (2) when the transition starts.

In this work, by a novel nanoscopic velocity measurement technique — laser induced fluorescent photobleaching anemometer (LIFPA)[18,20], we experimentally investigated the detailed fluctuation dynamics of a simple but fundamental EOF on an insulated wall driven by an AC electric field at sufficient high spatial-temporal resolution. The velocity fluctuations that are linearly proportional to the electric field in bulk fluids are precisely measured and ascribed to reveal the in-situ response of the flow field due to the external electric control in a wide range of magnitudes.

**Experiments setup**

The LIFPA system has simultaneously high temporal (~9.6 μs) and spatial resolutions (~180 nm and <800 nm in lateral and axial directions of the laser beam). The experiments is conducted in a microchip diagramed in Fig. 1a. We measured velocity fluctuations on the insulated bottom wall. The working fluid is coumarin 102 dye solution prepared with DI water and PBS buffer (see supplementary materials). Debye length $\lambda$ ( = $\sqrt{\varepsilon k_B T / N_A e^2 \sum_i c_i z_i^2}$ ) is estimated to be 42 nm. The AC field frequency $f_f$ ranges from 10 to 200 Hz, which is over $1/\tau_C$ (where $\tau_C = \lambda l/D = 0.42$ s is RC relaxation time), but smaller than $D/\lambda^2 \approx 2.83 \times 10^5$ Hz, indicating the diffusion layers on the electrodes are nonequilibrium but the diffuse layers stay in equilibrium. Therefore, the $f_f$ range is appropriate for studying

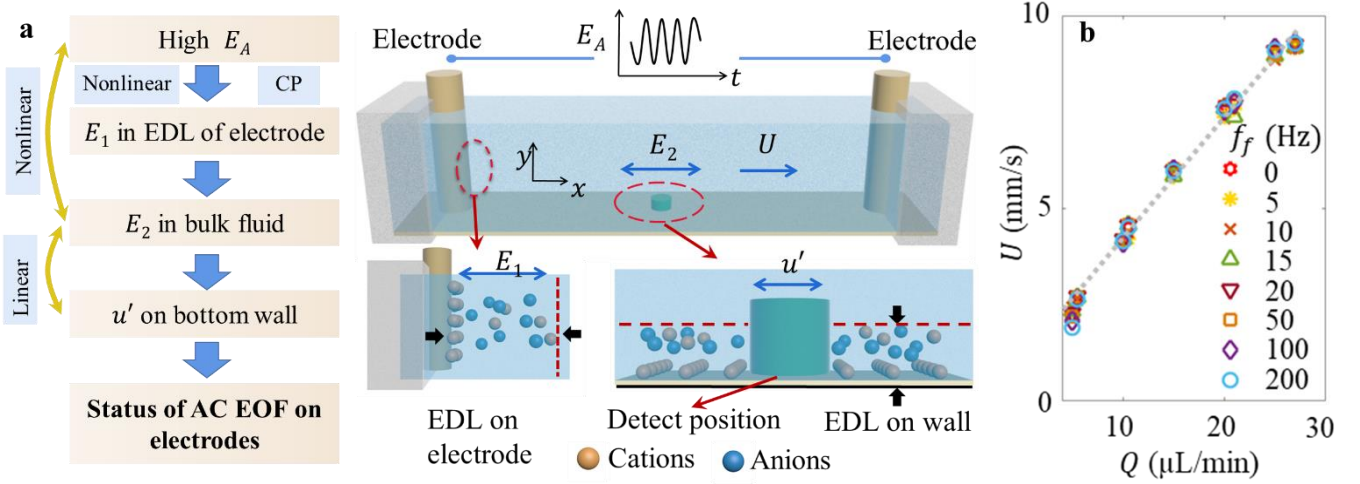

**Fig. 1 (a)** Diagram of AC EOF generated and its relation with the velocity fluctuation in microchannel. For linear AC EOF, the external electric field $E = E_A \sin(2\pi f_f t)$ has linear relationship with $E_1$, $E_2$ and $u'$. While for nonlinear AC EOF, the external electric field $E_A$ has nonlinear relationship with $E_1$ as a result of concentration polarization. Although $E_1$ has linear relationship with $E_2$ and $u'$ sequentially, $u'$ responds nonlinearly with $E_A$. Thus, we can distinguish the status of AC EOF through $u'$. In nonlinear status, $E_A$ and $u'$ has different topology of time trace or spectra information. Otherwise, the AC EOF is in linear status. **(b)** Mean velocity vs $Q$ with and without applying AC electric field measured at $z = 2$ μm, where $E_A = 8 \times 10^4$ V/m and $f_f$ =0-200 Hz.

nonlinear AC EOF where diffusion layer dynamics become dominant. Since the measurement position is at the center of the microchannel and 2.5 mm from each electrode, the local electric field related to externally applied AC electric voltage is approximately uniform and parallel (see Extended Data Fig. 7) in the streamwise direction. The problem we studied is cartooned in Fig. 1(a), i.e. externally applied AC voltage cause ion concentration variations around the electric double layers (EDLs) of electrodes, which in turn induce nonlinear AC electric field in the bulk region of microchannel flow, and subsequently, nonlinearly oscillating velocity field.

**Time series of velocity fluctuations and the statistics**

The measurement is taken at ~2 μm from the bottom wall, which is nearly two orders larger than the $\lambda$, but within the Stokes layer ($\geq 40$ μm) of the oscillating flow. It only takes $\tau_{md} = 4$ μs for the momentum to be transported to the measurement position where the largest velocity fluctuations along the vertical direction have been found. The momentum diffusion time $\tau_{md}$ is at least 3 orders smaller than $1/f_f$, which indicates the phase lagging of velocity time series at the measurement position relative to that in EDL of the insulated bottom wall is negligible. The mean velocities ($U$) at the measurement position with and without AC electric fields show no difference under $E_A = 8 \times 10^4$ V/m (which is the strongest electric field we applied in this investigation), and exhibit linear relations with flow rate ($Q$) (Fig. 1(b)). Thus, the basic flow has not been influenced by the applied AC electric field.

Fig. 2(a-f) show the influence of $E_A$ and $Q$ on the time series of $u'^* = u'/U_{HS}$, where $U_{HS} = \varepsilon|\zeta|E_A/\mu$ is Helmholtz-Smoulochowski velocity. In Fig. 2(a) where $f_f = 10$ Hz, as $E_A$ is increased, $u'^*$ behaves from sinusoidally to non-sinusoidally and asymmetrically with decreasing amplitude. Similar trends can also be found in Fig. 2(b, c) where $f_f$ =20 and 50 Hz. All these indicate increasing $E_A$ can enhance the nonlinearity of AC EOF. In contrast, the increasing $Q$ tends to inhibit the nonlinear AC EOF, as can be found from the decreasingly non-sinusoidal and asymmetric time series of $u'^*$ in Fig. 2(d-f). These findings are consistent with previous investigations [17] and support that $E_A$ and $Q$ are control parameters of AC EOF.

As the amplitude of the external AC electric field ($E_A$) increases, the velocity fluctuations evaluated by $u_{rms} = \sqrt{u'^2}$ ($u' = u - U$ is the velocity fluctuation, with $u$ being the instant velocity) can be clearly distinguished into two regions (Fig. 2(c)). $u_{rms}$ increases with $E_A$ linearly first, and then attain nonlinear saturation. The linear region extends with $U$. The slope and width of the linear region are also related to $f_f$. As plotted in Fig. 2(d), at lower $f_f$, the $u_{rms} \sim E_A$ plot has larger slopes and narrower linear regions.

In the nonlinear region, $u_{rms}$ is proportional to $E_A^b$, with $b = b(U)$ as plotted in the inset of Fig. 2(a). As $U$ is increased, $b$ increases towards unity. The nonlinear region turns into a linear one. By extrapolation, $b(0) \approx 0.5$, which is consistent with the observation of Levitan et al. [21] in induced-charge electroosmosis.

**Velocity power spectra**

Changing of the time series of $u'$ from sinusoidal to non-sinusoidal denotes a spectral distribution variation in frequency space. The evolution of AC EOF from linear to nonlinear can be more clearly distinguished by a sensitive spectral analysis from the power spectra of $u'$. An example is plotted in Fig. 3(a). At low $E_A = 5 \times 10^3$ V/m, there is only a single peak of $S(f)$ at 10 Hz which is also the forcing frequency. As $E_A$ is increased, accompanied by the increasing peak values of $S(f)$ at 10 Hz, the peaks at second, third and higher-order harmonic frequencies of $f_f$ gradually emerge, accompanied by increasingly nonlinearity.

We monitor the change of the peak $S(f)$ with increasing $E_A$ (Fig. 3(b)). When $E_A < E_{C2}$, the $S(f)$ values at the harmonic frequencies of $f_f$, i.e. $f/f_f =$2, 3, 4 and 5, are all on the noise level, i.e. no distinguishable peaks. When $E_A$ is increased beyond $E_{C2}$, a second peak emerges at the second harmonic frequency of $f_f$ ($f/f_f =$2). As $E_A$ is further increased, more peaks emerge at the third, fourth and fifth harmonic frequencies of $f_f$, if $E_A$ is larger than $E_{C3}$, $E_{C4}$ and $E_{C5}$ sequentially. Among the $E_{Ci}$ ($i =$ 2, 3, 4 and 5), $E_{C2}$ indicates the smallest $E_A$, above which the nonlinear AC EOF becomes observable. Therefore, as a technical criterion, we define $E_{C2}$ to be the experimentally valid and observable critical $E_A$ (say $E_{A,C}$) where the onset of nonlinear AC EOF happens.

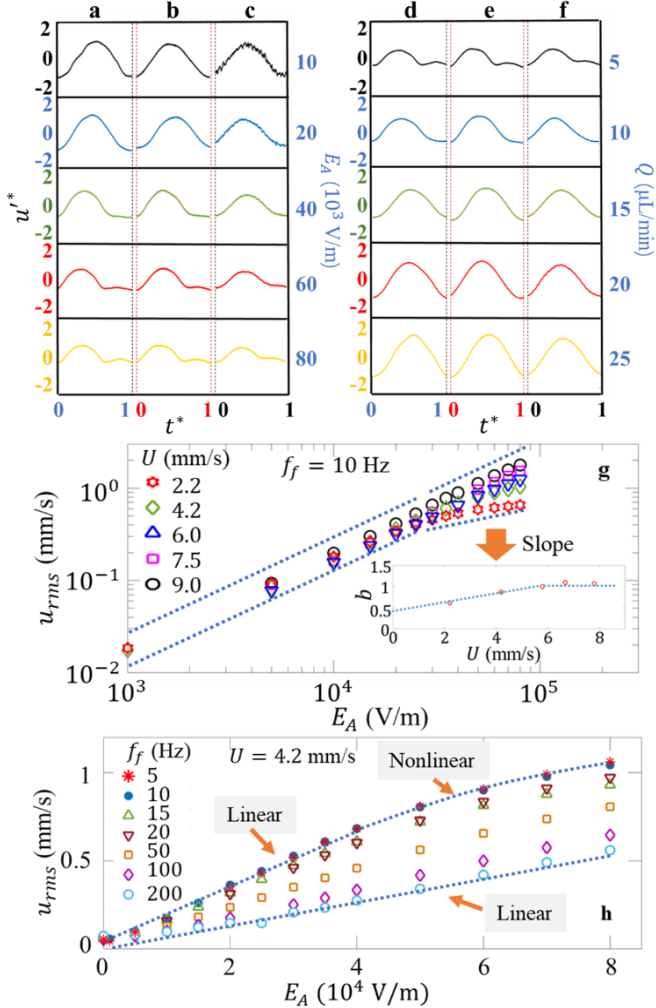

**FIG. 2 Phase-averaged $u'^*$ time series and the root-mean-square values $u_{rms}$ of velocity fluctuations. (a-c)** Phase-averaged $u'^*$ of $f_f =$10, 20 and 50 Hz respectively, at $Q =$10 μL/min. As $E_A$ is increased from top to bottom, the nonlinearity of AC EOF increases accompanied by increasingly non-sinusoidal $u'^*$. **(d-f)** Phase-averaged $u'^*$ of $f_f =$10, 20 and 50 Hz respectively, at $E_A = 8 \times 10^4$ V/m. As $Q$ is increased from top to bottom, the nonlinearity of AC EOF decreases accompanied by decreasing non-sinusoidal $u'^*$. **(g)** $u_{rms}$ vs $E_A$ at $f_f =$10 Hz and different $U$. The inset shows the slopes ($b$) of the log-log plots versus $U$ in the nonlinear region. In the log-log plot, when $b$ is increased from 0.5 to 1, the flow changes from nonlinear to linear status. **(h)** $u_{rms}$ vs $E_A$ at $U = 4.2$ mm/s and different $f_f$.

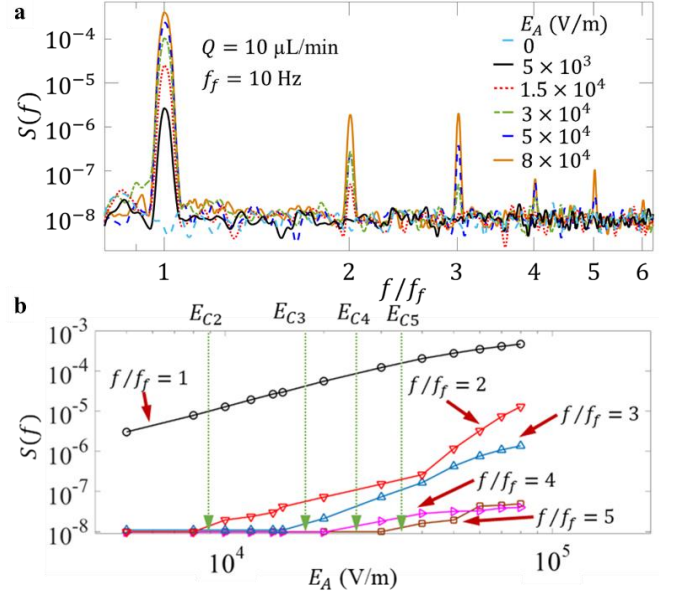

**Fig. 3. Power spectra of $u'$ and an illustration of the critical electric field. (a)** Change of velocity power spectra $S(f)$ with $E_A$. **(b)** Change of $S(f)$ peaks at different harmonic frequencies with $E_A$. $S(f)$ peaks at the first harmonic frequency ($f/f_f = 1$) increases with $E_A$ asymptotically. While the $S(f)$ peaks at the higher-order harmonic frequencies ($f/f_f =$2, 3, 4 and 5) are initially independent of $E_A$, they increase with increasing $E_A$ rapidly when $E_A$ is above a critical value, i.e. $E_{Ci}$ which is the critical $E_A$ beyond which the $i^{th}$ harmonic frequency emerges. $E_{Ci}$ increases with $i$. We define $E_{A,C} = E_{C2}$.

**Critical electric field intensity versus $U$ and $f_f$**

$E_{A,C}$ is not constant but proportional to $U$ and $f_f$, as plotted in Fig. 4(a) and (b). For instance, at $U =$2.2 mm/s and $f_f = 5$ Hz, dimensionless electric field $E^*_{A,C} = E_{A,C}\lambda/V_{ther} \approx 0.01$. However, at $U =$9.0 mm/s and $f_f = 200$ Hz, $E^*_{A,C} \approx 0.0665$. We estimated $E^*_{A,C}$ in quiescent fluid as $U \to 0$ by nonlinearly extrapolating the curves of Fig. 4(a), and calculated $E^*_{A,C}$ for the breakdown of weakly nonlinear AC EOF theoretically according to Eq. (70) of Olesen et al.[22]. As plotted in the inset of Fig. 4(a), both results of $E^*_{A,C}$ at $U = 0$ in experiments and theory show acceptable consistency.

The log-log plot of $E^*_{A,C}$ and $f_f$ exhibits a linear relationship (as inferred from Fig. 4(b)), with the slope directly related to $U$ as

$$E^*_{A,C} \sim f_f^{n(U)} \quad (1)$$

where $n(U)$ is linearly related to $U$ (inset of Fig. 4(b)), as

$$n(U) \approx -0.027U + 0.48 \quad (2)$$

Eq. (1) and (2) are one of the major findings of this investigation. They indicate there is a strong and complicated entanglement among $E_{A,C}$, $f_f$ and $U$. $E_{A,C}$ monotonically increases with $f_f$ and $U$. In other words, if a fixed and strong $E_A$ is applied (beyond a certain threshold value), nonlinear AC EOF can always be found when $f_f$ and $U$ are sufficiently small. When $U \rightarrow 0$, we have $E_{A,C} \sim f_f^{0.48}$. This is closely consistent with the investigation of Olesen et al. [23], who indicates the critical voltage for the breakdown of weakly nonlinear AC EOF is proportional to $f_f^{1/2}$, when $2\pi f_f \tau_C \gg 1$ and $\lambda/L \ll 1$.

The consistency between the experiments at $U \rightarrow 0$ and theory [24] is, to be honest, beyond our expectations. The experimental $E_{A,C}(U = 0)$ we are seeking is the threshold of weakly nonlinear AC EOF. While the theory [25] for comparison is about the breakdown of weakly nonlinear AC EOF. Surprisingly, the two supposed different status of AC EOF exhibits similar results. One possible explanation is the electrolyte solution we used is very dilute with relatively thick EDL. The crowding effect is negligible. Thus, the width of $E_A$ of the weakly nonlinear subregion is very narrow, which subsequently, is accompanied by a nearly overlapping of the threshold $E_A$ and the breakdown $E_A$ in the weakly nonlinear region.

**Universal dynamics parameters**

Determining the universal control parameters for distinguishing different EOF statuses from linear to nonlinear is always a top priority for theoreticians and engineers who are seeking a practical strategy to utilize nonlinear EOF. Former investigation [26] indicates the velocity fluctuation in an EOF driven by AC electric field is governed by an inviscid and general Burgers' equation with forcing term

$$\frac{\partial u'^*}{\partial t^*} + Z_l \frac{\partial u'^*}{\partial x^*} + Z_{nl} \frac{\partial u'^{*2}}{\partial x^*} = -2\pi \cos(2\pi t^*) \quad (3)$$

if a sinusoidal electric field is applied. The status of the flow field is determined by two dimensionless parameters, which are $Z_l = P_{e,U}/StSc\gamma^2$ and $Z_{nl} = P_{e,EP}/2StSc\gamma^2$, where $St = \lambda^2 \rho_f f_f / \mu$, $Re = \rho_f U_{HS} \lambda / \mu$, $Sc = \mu/\rho_f D$, $P_{e,EP} = z^2 e E_A L / k_B T$ and $P_{e,U} = UL/D$ are the Stokes number, electroosmotic Reynolds number, Schmidt number, Peclet numbers related to electrophoresis and local mean flow velocity respectively. $\gamma = L/\lambda$ is a length scale ratio between the outer scale and EDL scale. $Z_l$ and $Z_{nl}$ control the linearity and nonlinearity of the flow respectively. The former is directly related to $U$ and denotes the ratio of convective transports between the basic flow and oscillation, while the latter is changing with $E_A$ and denotes the ratio of electroconvection with respect to oscillation. We calculated all the critical $Z_{nl}$ (say $Z_{nl,C}$) related to $E_{A,C}$ against the $Z_l$, including in a broad parameter space of various $f_f$ and $U$. As shown in Fig. 4(c), all the data points fall on a single critical curve (red color), which can be expressed with $Z_{nl,C} \sim Z_l^{a_1}$, with $a_1 = 0.88$ Below the critical curve, the velocity field is linearly responding to the external AC electric field. No concentration polarization is needed to be considered. While above the critical curve, weakly nonlinearity of AC EOF becomes increasingly important. This is the second major finding of this investigation.

**Conclusions**

To the best of our knowledge, for the first time, the transition of AC EOF from linear to nonlinear has been experimentally studied directly relying on velocity measurement. A universal curve, $Z_l \sim Z_{nl,C}$, has been concluded and supports the validity of $Z_l$ and $Z_{nl}$ as universal control parameters. The nonlinear AC EOF has an important influence on broad fields, either positive or negative. In electrochemistry systems like supercapacitors and liquid electrolyte batteries, during fast charge and discharge, nonlinear AC EOF causes the loss of electric energy and should be inhibited, while in the microfluidics systems like micromixers and microreactors, nonlinear AC EOF can potentially enhance mixing and transport on the solid-fluid interface. Properly switching the optional linear and nonlinear AC EOF is crucial for different applications. The critical $Z_l \sim Z_{nl}$ curve provides an instructional guide for engineering applications and a solid foundation of theoretical analysis on nonlinear EOF subsequently.

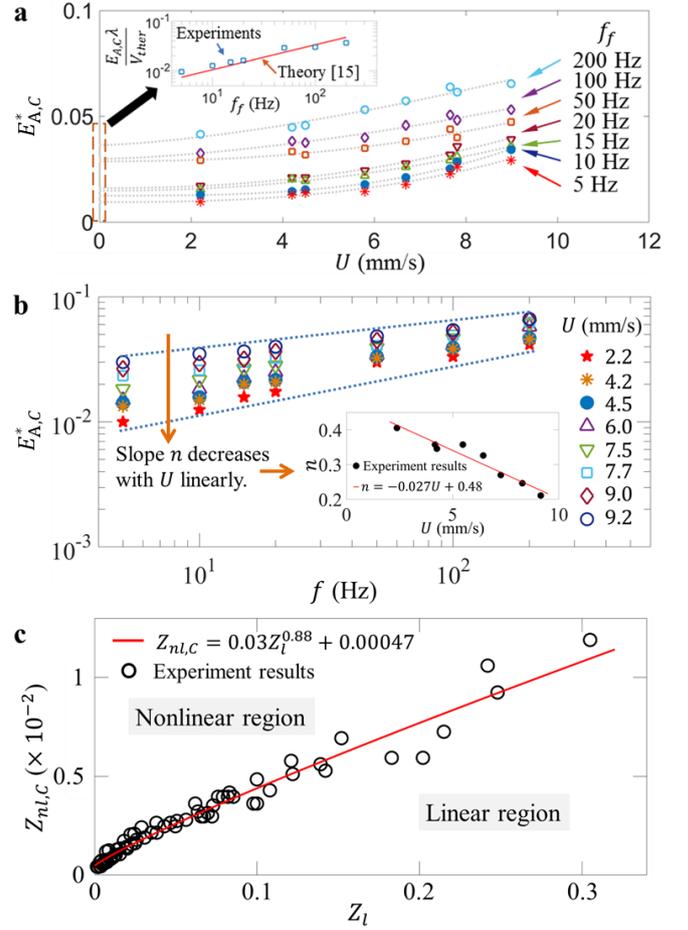

**Fig. 4** Dimensionless critical electric field $E_{A,C}^* = E_{A,C}\lambda/V_{ther}$ varies with $U$ at different $f_f$. $V_{ther} = k_B T/ze$ is thermal potential. **(a)** $E_{A,C}^*$ vs $U$ for different $f_f$. The inset shows $E_{A,C}^*$ vs $f_f$ at $U = 0$ from experiments by extrapolation and theoretically predicted breakdown electric field for weakly nonlinear AC EOF [15]. **(b)** $E_{A,C}^*$ vs $f_f$ for different $U$. Each curve can be expressed by Eq. (1) with a power of $n(U)$, which is also the slope of the log-log plot. $n(U)$ is fitted by Eq. (2), as shown in the inset. **(c)** Universally critical curve $Z_l \sim Z_{nl,C}$ between linear and nonlinear AC EOF.

**Acknowledgement**

The investigation is supported by National Natural Science Foundation (No. 11672229, 51927804, 61775181, 61378083), the Natural Science Basic Research Program of Shaanxi Province (2016ZDJC-15, S2018-ZC-TD-0061, 2018JM1061).

## Methods

**Laser induced fluorescence photobleaching anemometer (LIFPA) technique**

LIFPA is a micro/nanofluidics velocity measurement technique depending on fluorescent photobleaching. During LIFPA application, the working fluid is uniformly mixed with fluorescent dye. When the working fluid is irradiated by laser, the fluorescent intensity of the fluorescent dye will rapidly decrease with time, as a result of photobleaching. When the concentration of the fluorescent dye is constant, the longer it is irradiated by the laser, the stronger the photobleaching effect and the weaker the fluorescence intensity that can be collected. This process can be approximately described by an exponential equation as [23]

$$I_f = I_{fo} e^{-t/\tau} \quad (4)$$

where $I_{f0}$ represents the fluorescence intensity at the initial time, $t$ represents the resident time under irradiation, and $\tau$ represents the photobleaching time constant. When the solution containing fluorescent molecules passes through the laser spot, if the velocity is $u$ and the focus diameter of the laser beam is $d_f$, the resident time for the fluorescent molecules to be irradiated within the laser focus can be described by the convection velocity and spatial position, as

$$t = x/u \quad (5)$$

with $x \in [0, d_f]$. Combining equations (4) and (5), we have

$$I_f = I_{fo} e^{-d_f/u\tau} \quad (6)$$

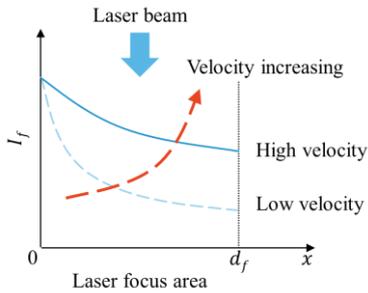

**Extended Data Fig. 1** Principle of LIFPA. Under the irradiation of laser, the fluorescence intensity decreases along the flow direction (e.g. x-direction in this research) in the focus of the laser beam.

Equation (6) is a simplified model of the photobleaching process. It can be found from this equation that when $I_{f0}$ (the concentration of fluorescent dye) is constant, the faster the fluid velocity, the shorter the time it stays in the laser focal spot area. The level of photobleaching is accordingly lower (see **Extended Data Fig. 1**), with more fluorescent intensity $I_f$ collected in the focal region of the objective lens. Therefore, a larger velocity magnitude of the fluid is monotonically accompanied with stronger fluorescent intensity we can measure, as can be seen later from the LIFPA calibration curve. By measuring fluorescent intensity, the magnitude of flow velocity can be determined.

LIFPA is non-invasive optical method with several fantastic features, e.g. simultaneously high spatiotemporal resolution, no interference from electric fields, and the ability to perform far-field nanoscale velocity measurement. It has been widely used in the investigations on the velocity fields of micro electrokinetic turbulence[27-29] and electroosmotic flow etc[14,25]. The temporal resolution ($t_s$) of LIFPA is dominated by the photobleaching time constant $\tau$ as[27]

$$t_s \geq 2\pi\tau \left(\frac{U^2}{u_a^2} - 1\right)^{-1/2} \quad (7)$$

where $U$ is the average velocity and $u_a$ is the amplitude of the velocity fluctuation. From equation (7), $t_s$ is linearly related to $\tau$. Meanwhile, it can also be found that the larger $U/u_a$, the smaller $t_s$, and accordingly the higher the time resolution. However, even if $u_a$ approaches $U$, e.g. $u_a = 0.95U$ (which is far from achievable in this investigation), we have $t_s \approx 6\pi\tau$. Considering $\tau$ itself is 9.6 μs in this investigation, $t_s$ is below 200 μs which is sufficiently high to measure the oscillating velocity of AC EOFs. The spatial resolution is determined by the optical system used and will be introduced in detail below.

**Experimental setup**

The LIFPA system used in the experiment was developed on the basis of a confocal microscope system, as shown in **Extended Data Fig. 2**. This system uses a continuous laser (MDL-III-405-500) with a wavelength of 405 nm, and the maximum power of the laser is 500mW. At the output end of the laser, an acousto-optic modulator (AOM, 1206C-2-1002, Isomet) was first used to modulate the beam temporally. Its aperture is 2 mm. The first-order diffraction spot is selected as the excitation beam. The quality of the light beam passing through the AOM will degrade so that a spatial light filter (SLF, SFB-16DM, OptoSigma) is applied at downstream to obtain a standard Gaussian spot, and then. collimated by a lens. The light beam enters the dichroic mirror through the mirror. The purpose of the dichroic mirror we choose here is to separate the laser from the fluorescent signal. Therefore, the light beam will be reflected by the dichroic mirror into the objective lens, and then focus on the sample. A 100 × NA 1.4 oil immersion objective lens (Olympus PlanApo) was used in the experiment. The fluorescent signal of dye excited by the laser passes through the objective lens and the dichroic mirror in turn. Before the fluorescent signal is collected by a multimode fiber (25 μm in diameter, 400-550 nm band, Thorlabs, M67L01) and detected by a single photon counter (Hamamatsu H7421), two OD4 bandpass filters (470/100 nm and 470/10 nm) are used to ensure the block of nonfluorescent signals.

In order to ensure the positioning accuracy within a large displacement range, two translation stages are used at the same time, including a 2D electric translation stage (Physik Instrumente, PI, M-521.DG) with 1 μm precision over 300 mm travel distance, and a high-precision 3D nano translation stage (Physik Instrumente, PI, 562.3CD) with 1 nm precision over 200 μm travel distance. In the experiments, a self-developed control system with software was used to control the system and count the number of fluorescence photons transmitted by photon counter in real time. The lateral and axial spatial resolutions of the system are approximately 180 nm (as can be seen from **Extended Data Fig. 2**(b) and (c)) and 800 nm, respectively.

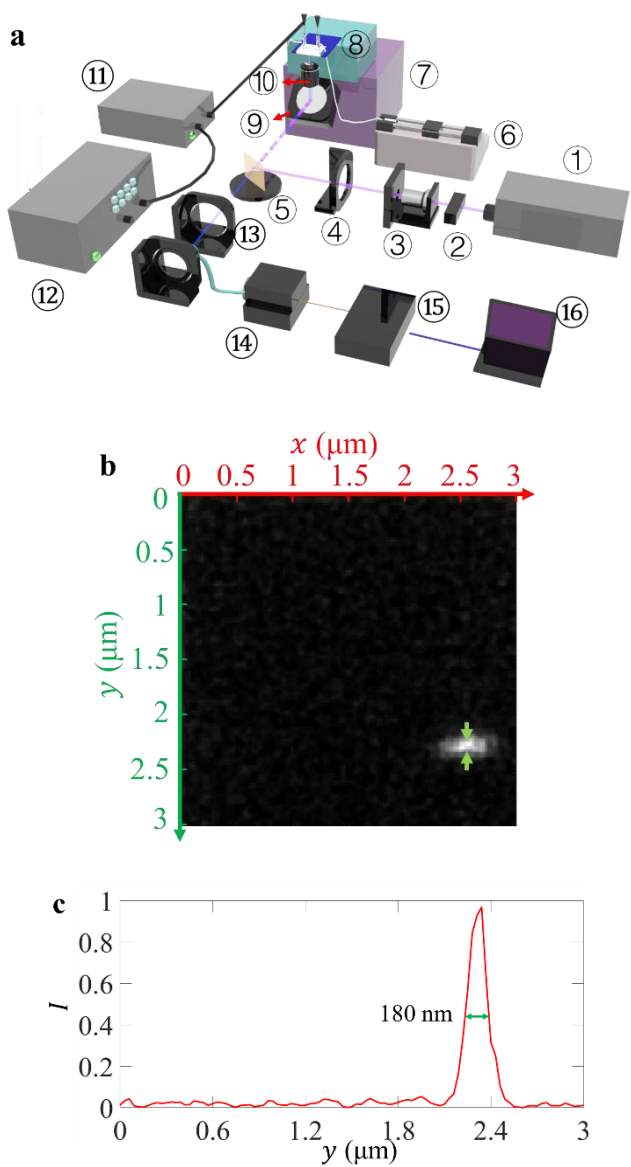

**Extended Data Fig. 2** Schematic diagram of LIFPA system and its spatial resolution. (a) Schematic diagram of LIFPA system, including: (1) 405 nm continuous wave laser; (2) acousto-optic modulator (AOM); (3) spatial light filter; (4) lens; (5) dichroic mirror; (6) syringe pump; (7) 2D translation stage with micro resolution; (8) 3D piezo translation stage with nano resolution; (9) mirror; (10) 100X oil immersion objective lens; (11) voltage amplifier; (12) function generator; (13) bandpass filter set (one 470/100 nm and one 470/10 nm); (14) photon counter; (15) control box; (16) computer. (b) Schematic diagram of scanning a gold particle (20-25 nm diameter) with the confocal system to demonstrate the spatial resolution of LIFPA system. (c) Light intensity distribution along the lateral direction. The full width at half maxima of the profile, i.e. 180 nm, represents the lateral spatial resolution of the confocal and LIFPA system.

In the experiments, we used a rectangular microchannel, whose length ($l$), width ($w$) and height ($h$) are 5 mm, 300 μm and 100 μm, respectively (**Extended Data Fig. 3**). The chip consists of three layers, which are processed layer-by-layer. The cover layer is made of acrylic sheet with a thickness of 2 mm. It has good light transparency and mechanical strength, which can fix the whole chip well. The middle layer is a channel layer made of plastic sheet. The bottom layer is a glass slide having a good ultraviolet transmittance and a thickness of 130 μm. The entrance and exit of the channel are T-shaped plastic joints. Platinum electrodes with diameter of 100 μm are inserted into the entrance and exit, and sealed. The solution in the channel is provided by a pressure-driven syringe pump (HARVARD Apparatus PUMP 33) through 1 mL syringe. PEEK tubing is applied to connect the syringe and the T-joint. Here, as shown in **Extended Data Fig. 3**, a three-way valve is used between the syringe and the T-joint. This is to keep the entire microchip from entering bubbles, and degas if necessary. In the experiment, an AC electric field is applied on the electrodes. The AC electric field is provided by a function generator (Tektronix AFG3102C) and a high-voltage amplifier (Trek PZD700A). At the same time, the high-voltage amplifier needs to be grounded to guarantee precise amplification.

In the LIFPA measurement, 100 μM Coumarin 102 aqueous solution is used. Its excitation and emission spectra are plotted in **Extended Data Fig. 3**(b). The pH value and electric conductivity are 7.1 and 6.1 μS/cm, respectively. The concentration of methanol is 5%.

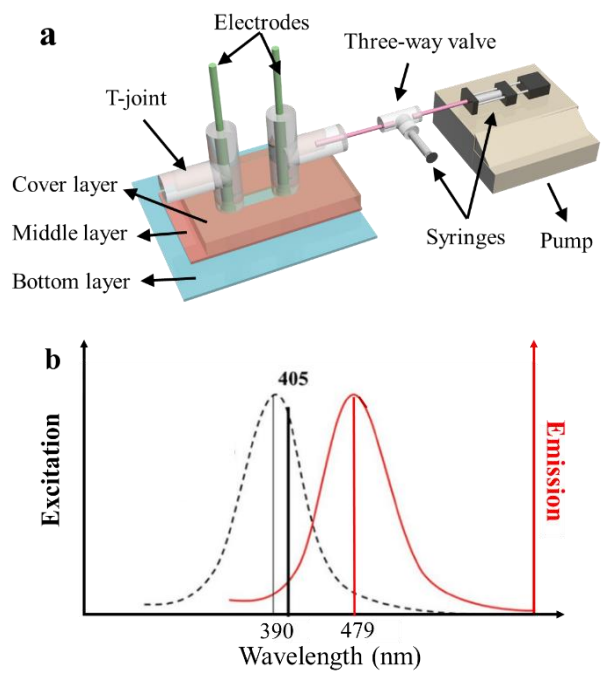

**Extended Data Fig. 3** (a) Schematic of the microchip. (b) Excitation and emission spectra of Coumarin 102 (from Sigma-Aldrich).

**Fluorescent signal measured in the LIFPA system**

In the LIFPA system, the detector is a photon counter which is designed for extra low light detection. It can be easily saturated when the fluorescent signal is slightly increased. Therefore, an OD3 neutral density filter was applied to reduce fluorescent signal. The fluorescent signal is captured in a sequence shown in **Extended Data Fig. 4**.

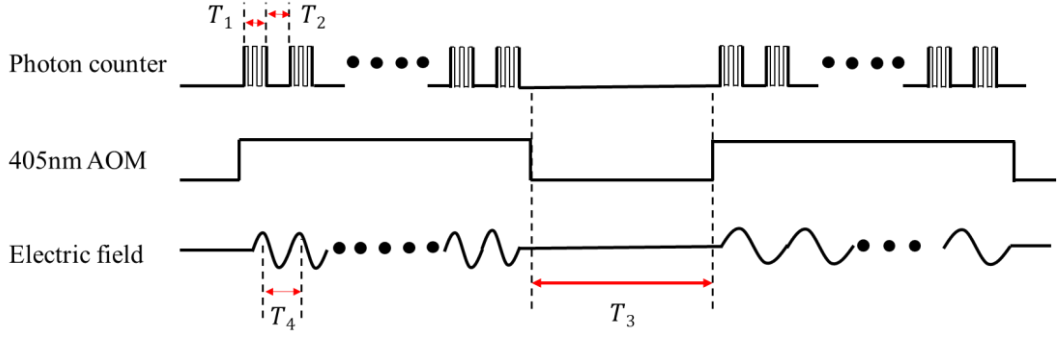

**Extended Data Fig. 4** The sequence of the measurement signal. $T_1 = 500$ μs is exposure time of photon counter, $T_2 = 500$ μs is delay time, $T_3 = 5$ min is the interval between two measurements. $T_4 = 1/f_f$ is the period of external electric field. $T_4$ can be different between two measurements.

In the experiment, we give the AOM a square wave signal to control the on/off of the beam, then, synchronize the photon counter to measure fluorescence. The exposure time $T_1$ for photon counter is 500 μs. Since the rise time of AOM is usually less than 1 μs, the delay ($T_2$) of 500 μs is sufficiently long for the laser power output from AOM to be stabilized. $T_3$ is the interval between two sets of experiments. In order to ensure the stability of the experimental system, we set the interval to 5 minutes. At the same time, the period $T_4$ of the AC electric field in each set of experiments is determined by $1/f_f$ given by the function generator. Thus, we can effectively collect the time series of fluorescence signals corresponding to velocity fluctuation of AC EOF.

**LIFPA calibration**

During the experiment, the first step is to measure the velocity calibration curve. The velocity calibration is carried out at the center of the microchannel, where the fluorescent signal is the largest in the cross section. As shown in **Extended Data Fig. 5**, the initial position of the laser focus may not be in the center of the channel. Therefore, we need to determine the center position of the channel by moving the position of the channel and monitoring the number of fluorescent photons. After finding the center of the channel, the velocity calibration curve in the microchannel can be measured. Giving a series of flow rate $Q$, the corresponding flow velocity ($U$) at the center of microchannel can be calculated by [30]

$$U(y,z) = \frac{48Q}{\pi^3 wh}\left\{\sum_{n,odd}^{\infty}\frac{1}{n^3}\left[1-\frac{\cosh\left(\frac{n\pi y}{h}\right)}{\cosh\left(\frac{n\pi w}{2h}\right)}\right]\sin\left(\frac{n\pi z}{h}\right)\right\}$$
$$\left[1-\sum_{n,odd}^{\infty}\frac{192h}{n^5\pi^5 w}\tanh(\frac{n\pi w}{2h})\right]^{-1} \quad (8)$$

In the meanwhile, the fluorescent signal is detected by photon counter. Subsequently, a monotonic relation between fluorescent signal and flow velocity can be established, as shown in **Extended Data Fig. 5**(b). The $U(0,0)$~$I_N$ curve is nonlinearly fitted by a polynomial function plotted with black line. The fitting is perfectly consistent with the experimental results. Finally, the polynomial function is used to calculate $u$ at the measurement position.

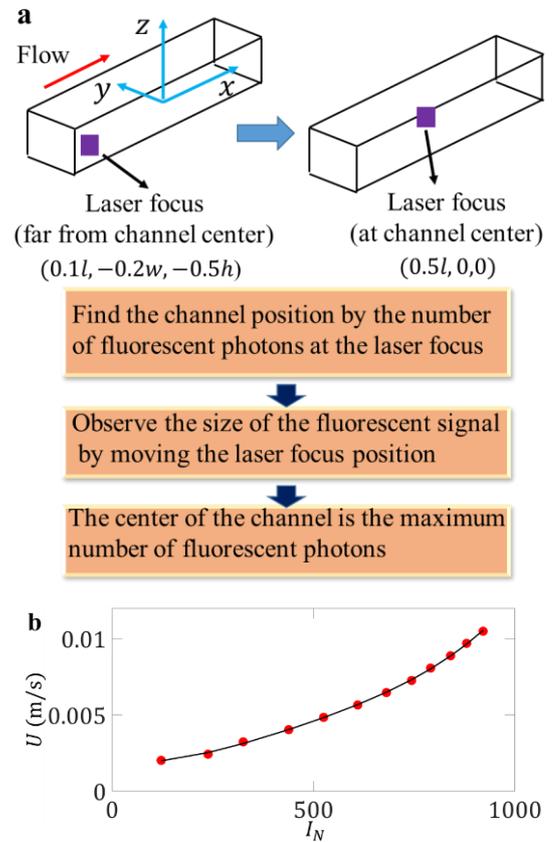

**Extended Data Fig. 5** Procedure of determining the center of microchannel for velocity calibration and a typical velocity calibration curve. (a) When the laser focus was in the microchannel, the number of fluorescent photons could be measured. At this time, the laser focus position was probably not located at the center of the microchannel. Therefore, we monitored the number of the fluorescent photons along with moving the position of the microchannel, when the photon number reaches the maximum, the laser focus located at the center of the channel. (b) Velocity calibration curve at the center of microchannel. The excitation beam power at the pupil of objective lens is 11 mW.

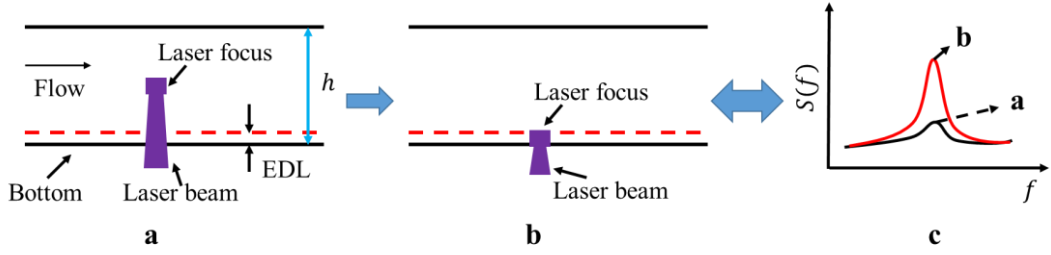

**Extended Data Fig. 6** Procedure of determining the measurement position around EDL and its relationship with the peak value of velocity power spectrum ($S_f$). (a) Laser focus is far away from EDL. (b) Laser focus is close to the EDL where the power spectra peak of the velocity fluctuation reaches maxima. In this investigation, $z_m \approx 2$ μm. (c) Power spectra peak of the velocity fluctuation at $f_f$ varying with different $z$ positions according to (a) and (b).

**Determination of the measurement position**

After the velocity calibration is completed, it is necessary to find the measurement position for this experiment—the electric double layer (EDL) near the insulated wall. At this time, we set $Q = 10$ μL/min, and applied an AC electric field with a frequency ($f_f$) of 10 Hz and a low voltage ($V_{ext}$) of 100 Vpp (which is close to but below the critical voltage of linear-nonlinear AC EOF) across the channel. Use the nano translation stage to move the microchannel upwards (**Extended Data Fig. 6**(a)) and monitor the fluctuation of the photon number. When the photon number fluctuates near the maximum value, we calculate the magnitude of the power spectrum peak of the velocity fluctuation at 10 Hz. By fine-tuning the microchannel position, we can find a peak where the power spectra of the velocity fluctuation at 10 Hz reach maxima (**Extended Data Fig. 6**(c)). Here is the position of the EDL on the bottom surface (**Extended Data Fig. 6**(b)), which is the position where the velocity fluctuation reaches the largest. In this investigation, the largest velocity fluctuation appears at $z_m \approx 2$ μm. Although this position is not on EDL, the oscillation of EDL can be transported to the measurement position in $z_m^2 \rho_f/\mu \approx 0.1$ ms. Since the investigated frequency range is below 500 Hz (from spectra), all these spectral components can be transported to the measurement position in time. Therefore, at $z_m$, we can still measure oscillating velocity with sufficiently rapid response.

**Reliability of LIFPA**

Briefly, $2\pi u_{rms} f_f$ which owns the dimension of acceleration qualitatively determines the fastest response of LIFPA. From Fig. 2(b) of the manuscript, the maximum of $2\pi u_{rms} f_f$ in the linear region is no less than 0.7 m/s². In other words, as an example, LIFPA here can measure 10 Hz oscillating velocity up to 10 mm/s with reliable resolution. All the $u'$ measured in this investigation have $2\pi u_{rms} f_f \leq 0.7$ m/s², as can be estimated from Fig. 2(a) and (b) of the manuscript. This supports the reliability of velocity measurement in this investigation.

**Estimation of critical electric field at $U = 0$**

In the investigation of Olesen et al. [29], they provided a theoretical prediction (Eq. 70 in their paper) on the breakdown voltage of weakly nonlinear AC EOF, in a quiescent fluid. Here, on the basis of our experimental conditions, we calculate the corresponding critical electric field ($E_{A,C}$ ($U = 0$) $= V_C/L$) and compared with their theory. The results are summarized in Table 1.

For comparison, we also estimated the $E_{A,C}$ at $U = 0$ by extrapolating our experimental data, as displayed in **Extended Data Fig. 5**(a) of the manuscript. The extrapolated $E_{A,C}$ is also shown in Table .

Table 1. Comparison of $E_{A,C}$ ($U = 0$) (V/m) from this experiment and the theory of Olesen et al. [15]

| $f_f$ (Hz) | 5 | 10 | 15 | 20 | 50 | 100 | 200 |
|---|---|---|---|---|---|---|---|
| Experiment | 5881 | 7745 | 9161 | 9892 | 17650 | 18360 | 22340 |
| Theory [25] | 4550 | 6435 | 7881 | 9100 | 14388 | 20348 | 28776 |

* Conditions for calculation: $\varepsilon = 7.1 \times 10^{-10}$ F/m, $z = 1, e = 1.602 \times 10^{-19}$ C, $k_B = 1.38 \times 10^{-23}$ J/K and $T = 298$ K, $\lambda = 42$ nm, Stern layer thickness $\lambda_S = 0.7$ nm [31], $L = 5$ mm, effective diffusivity $D = 5 \times 10^{-10}$ m²/s.

**Electric field distribution**

In **Extended Data Fig. 7**, we plot the distribution of the electric field and electric potential, which are simulated by COMSOL Multiphysics 4.3. It can be seen, at the position of velocity measurement, the electric field is parallel to the streamwise direction of the microchannel, with uniform intensity. This supports our approximation in the **Experimental Setup** section.

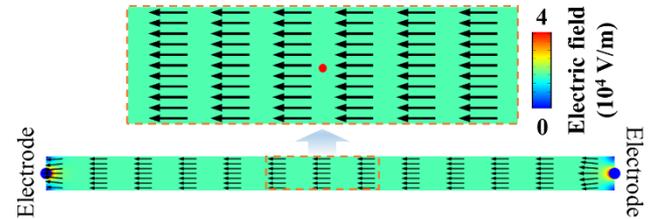

**Extended Data Fig. 7** Electric field distribution. The arrows show the direction of the electric field. The colormap shows the magnitude of the electric field. The red dot indicates the position of velocity measurement in AC EOF. The red box indicates the local zoom-in of the field. Here, the influence of concentration polarization was not taken into account.

**Dimensional and dimensionless parameters**

In the investigation, we have applied the dimenionless parameters as follows: $t = t^*/f_f$, $x = x^*L$, $u' = u'^*U_{HS}$, $E = E^*E_A$. The detailed experiemntal parameters have been listed in Table 2.

Table 2. Some parameters in experiments

| | |
|---|---|
| T (K) | 298 |
| $e$ (c) | $1.602 \times 10^{-19}$ |
| $k_B$ (J/K) | $1.38 \times 10^{-23}$ |
| $\varepsilon$ (F/m) | $7.1 \times 10^{-10}$ |
| $c_{H^+}$ (mol/L) | $1 \times 10^{-7}$ |
| $c_{OH^-}$ (mol/L) | $1 \times 10^{-7}$ |
| $c_{Na^+}$ (mol/L) | $5.15 \times 10^{-5}$ |
| $c_{Cl^-}$ (mol/L) | $5.15 \times 10^{-5}$ |
| Concentration of coumarin 102 (μmol/L) | 100 |
| $\zeta$ (mV) | -36 |
| $\mu$ (kg/m·s) | $10^{-3}$ |
| $\rho_f$ (kg/m³) | $10^3$ |
| $D$ (m²/s) | $5 \times 10^{-10}$ |
| $\lambda$ (nm) | 42 |